\documentclass[%
 reprint,
showpacs,
 amsmath,amssymb,
 aps,
]{revtex4-1}

\usepackage{graphicx}% Include figure files
\usepackage{dcolumn}% Align table columns on decimal point
\usepackage{bm}% bold math
\usepackage{xcolor}
\definecolor{blue}{rgb}{0,0.5,1.0}
\usepackage[export]{adjustbox}
\usepackage{ulem}
\begin{document}

\title{Lane formation in a driven attractive fluid}

%Authors
\author{C. W. W\"achtler}
\author{F. Kogler}
\author{S. H. L. Klapp}
\affiliation{
Institute of Theoretical Physics, \\
Secr. EW 7-1, Technical University Berlin, \\
Hardenbergstr. 36, D-10623 Berlin, Germany
}
\date{\today}

%%******Abstract*****%%
\begin{abstract} 
We investigate non-equilibrium lane formation in a generic model of a fluid with attractive interactions, that is, a two-dimensional 
Lennard-Jones (LJ) fluid composed of two particle species driven in opposite directions. 
Performing Brownian Dynamics (BD) simulations for a wide range of 
parameters, supplemented by a stability analysis based on dynamical 
density functional theory (DDFT), we identify generic features of lane formation in presence of attraction, including structural properties.
In fact, we find a variety of states (as compared to purely repulsive systems), 
as well as a close relation between laning and long wavelength instabilities of the homogeneous phase such as demixing and condensation.
\end{abstract}

%\pacs{05.65.+b}%{Self-organization in statistical physics}
%{Phase separation and segregation in colloids}
%\keywords{}

\maketitle

%%******Introduction*****%%
\section{\label{sec:Introduction}Introduction}
Lane formation is a prototype of a non-equilibrium self-organization process, 
where an originally homogenous mixture of particles (or other types of "agents") moving in opposite directions
segregates into macroscopic lanes composed of different species. 
This ubiquitous phenomenon may occur, e.g., in driven binary mixtures
of colloidal particles~\cite{leunissen2005,loewen:exp,loewen:first}
and migrating macro-ions~\cite{conduction:lane}, in binary plasmas~\cite{plasma:lane2,Suetterlin2009}, 
but also in systems of "active" (self-propelling) particles with aligned velocities such as bacteria in channels~\cite{lane:unidirectional} and humans in pedestrian zones~\cite{Helbing2000,Pedes}. 

Theoretically, lane formation has been studied extensively in model systems composed of hard or soft spheres, where the pair interactions are solely repulsive (see, e.g., \cite{finite_size:lane,Geissler2016}).
A realistic example are suspensions of charged colloids. For such systems, laning has been investigated concerning the impact of density~\cite{loewen:DFT-phasediagram},
the role of hydrodynamic interactions~\cite{loewen:hydro}, 
and the accompanying microscopic dynamics (particularly, the so-called dynamical locking)~\cite{loewen:exp}. More recently studied issues
are the impact of anisotropic friction \cite{attraction:lane} and of environment--dependent diffusion~\cite{Geissler2016}.

Compared to these repulsive (or predominantly repulsive) systems, lane formation in {\it attractive} systems has received much less attention. This contrasts the fact that
attractive interactions in colloids are quite common, ranging from isotropic (depletion or van-der-Waals) interactions to anisotropic ones. Examples for the latter are
the dipolar interactions between colloids with (permanent or induced) magnetic or electric dipoles, or the medium-generated interactions between colloids embedded in liquid crystals~\cite{schoenLCanisotropicinteraction}.
From an equilibrium perspective it is well established that such attractive (isotropic or anisotropic) interactions in colloidal systems can drive phase transitions, including condensation and demixing.
Moreover, particles with attractive forces, can be set into motion~\cite{Klappreview}, either by external fields (see, e.g., Ref.~\cite{gangwal:active}) or by intrinsic self-propulsion mechanisms~\cite{propulsionexp}, opening the possibility
for lane formation and related dynamical phenomena. Examples of swimmers with anisotropic (e.g., magnetic) interactions have been studied in~\cite{magneticswimmer1, magneticswimmer2, magneticswimmer3,review:janus2}.
 
In a recent study we have analyzed
the occurrence of laning in a system of oppositely driven dipolar microswimmers~\cite{kogler1} inspired by dielectric Janus particles~\cite{gangwal:active}. This study indicates that 
even strongly anisotropic attractive interactions have a profound impact on lane formation. In particular, laning was found to be correlated to a condensation phase transition of the underlying
{\it equilibrium} system. 
 
Motivated by these findings, we here present a systematic study of laning in a simple model involving only {\it isotropic} attractive interactions,
that is, a binary Lennard-Jones (LJ) fluid where particle species are driven against each other. The equilibrium LJ fluid is a prototypical system
exhibiting a condensation (gas-liquid) phase transition at sufficiently large strength of attraction. Further, its binary counterpart displays coupled condensation and demixing transition~\cite{Wilding1998}.
Thus, LJ models are ideally suited to identify generic features of laning in system with attractive pair interactions. 

Moreover, LJ interactions have already been shown to significantly affect the phase behavior of {\it active} particles (which, contrary to the particles considered in~\cite{kogler1}, can swim in any direction
due to rotational diffusion)~\cite{nonequi:review1}. For example, they dramatically change 
the phase behavior of active hard spheres~\cite{Redner2013_1,Redner2013_2} (which undergo phase separation due to self-trapping~\cite{Bechinger2013, activenucleation}), including reentrance of the homogeneous phase. 
 
The impact of LJ interactions on the laning transition of oppositely driven particles
has already been touched by some of us~\cite{kogler1} in an earlier simulation study based on Brownian Dynamics (BD); however, there we only considered one specific density.
Here we present BD results for a much larger parameter space (particularly a larger range of densities) combined with a detailed structural analysis. 
Furthermore, we supplement the simulations by a stability analysis based on Dynamical Density Functional theory (DDFT)~\cite{DDFTbasics, loewen:DFT-phasediagram}.
To this end we define an effective (equilibrium) model system in which the original pair interactions are corrected by the driving force. 
Similar attempts to map an intrinsically  non-equilibrum system onto an effective equilibrium one have been done in the context of active particles,
involving effective pair potentials~\cite{loewen:DFT-phasediagram,Schwarz-Linek,effective:active,loewenagainsteffective,Pedes}, an effective Cahn-Hillard equation~\cite{CahnHillardeff} or 
non-equilibrium equations of states~\cite{EQSSNEQ}.

Our results show that lane formation is indeed tightly related to the occurrence of long-wavelength instabilities (condensation and demixing) 
of the homogeneous phase of the {\it effective} model. Moreover, compared to purely repulsive systems we find a larger variety of non-equilibrium states with significant differences in their structure.

The rest of this paper is organized as follows. In Sec.~II we introduce the model and the methods of investigation. We also define the effective interaction used for the stability analysis.
In Sec.~III we start by giving an overview of the BD results, followed by a detailed analysis of the structural properties of the non-equilibrium states (Sec.~IIB). The last part of the section (Sec.~IIC) is then devoted
to a comparison of the BD results with those from the linear stability analysis. The paper closes with a brief summary and outlook.

%%******Model and Methods*****%%
\section{\label{sec:Model}Model and methods }
\subsection{\label{subsec:Simulation} Brownian Dynamics simulations}
We perform overdamped Brownian Dynamics simulations in a 2D quadratic cell of size $L^2$ with periodic boundary conditions along the $x$- and $y$-direction of the coordinate system. 
The cell contains up to $N=10000$  spherical particles of diameter $\sigma$, which serves as a unit length.
Each particle is assigned randomly a fixed number $s_i=1$ or $s_i=-1$ defining its species. Thereby, a 50:50 binary mixture is created.
The difference between the species is solely determined via the direction of the driving force $\mathbf f_{\text{d}}(s_i) = f_\text{d} s_i \mathbf e_y$, where $\mathbf e_y$ is a unit vector along the $y$-direction of the coordinate system. The driving force pushes the particle either
'upwards' ($s_i=+1$), i.e. along the y-axis, or 'downwards' ($s_i=-1$).

Independent of $s_i$, all particles interact via the Lennard-Jones potential
\begin{equation}\label{eqn:LennardJones}
U_{\text{LJ}} (r_{ij})= 4\varepsilon \left(\left(\frac{\sigma}{r_{ij}}\right)^{12}-\left(\frac{\sigma}{r_{ij}}\right)^6\right)~, 
\end{equation}
where $ r_{ij} = |\mathbf r_{ij}|=|\mathbf r_i - \mathbf r_j|$ denotes the distance between particles $i$ and $j$ and 
$\varepsilon $  is the interaction strength.
The potential is truncated at $r^\text{c}_{ij}=2.5\sigma$.\\ 
The equation of motion of particle $i$ is given by 
\begin{equation}\label{eqn:Lagevin}
\dot{\mathbf{r}}_i=\frac{1}{\gamma}\left[ \sum\limits_{j=1, j\neq i}^{N} -\boldsymbol\nabla U(r_{ij}) +\mathbf{f}_{\text{d}}(s_i) \right] +\sqrt{2D}\boldsymbol\xi ~,
\end{equation}
with the vector $\boldsymbol \xi$ representing Gaussian white noise. The Cartesian components of $\boldsymbol \xi$ fulfill the relations $\left<\xi_\alpha(t)\right> = 0$ and $\left<\xi_\alpha(t)\xi_\beta (t')\right>=\delta_{\alpha\beta}\delta(t-t')$ for $\alpha,\beta \in  \{x,y\}$. Here, $D=k_\text{B}T/\gamma$ is the diffusion constant with $k_\text{B}$ being Boltzmann's constant, $T$ the temperature and $\gamma$ the friction constant. 

In order to quantify the degree of laning along the driving force (i.e. the $y$-direction), we divide the simulation box into $k=1,..,K$ equal slices of 
widths $\sigma$ (along the $x$-direction) and length $L$ (along $y$-direction). 
For the $N_k$ particles in slice $k$, we sum up the orientations $s_i$. The ensemble and time-averaged laning order parameter is then defined as~\cite{attraction:lane}
\begin{equation}\label{eqn:OrderParamter}
\phi = \left< \frac{1}{K}\sum\limits_{k=1}^{K}\left|\sum\limits_{j=1}^{N_k} s_j/N_k\right|\right>.
\end{equation}
A non-laned system is characterized by $\phi\approx 0$ whereas a perfectly laned state corresponds to $\phi = 1$.

To characterize the local structure in $x$-direction we calculate the radial distribution function between particles of the same type
\begin{equation}\label{eqn:RadialDistributionX}
g^s(x) = \frac{2}{\varrho N} \left< \sum\limits_{i=1}^N \sum\limits_{j\neq i}^N \delta(x-|x_{ij}|)\Theta(\sigma-|y_{ij}|)\Theta(s_is_j)\right>
\end{equation}
with $x_{ij}$ ($y_{ij}$) being the $x$ ($y$)-component of $\mathbf r_{ij}$, and $\delta$ and $\Theta$ being the delta- and the Heavyside-step-function, respectively. 
By normalization, the function $g^s(x)$ decays to 1 for $x \rightarrow \infty$ in a perfectly mixed 50:50 binary system. 
The radial distribution function in $y$-direction, $g^s(y)$, is then defined by interchanging $x$ and $y$ in Eq.~(\ref{eqn:RadialDistributionX}). 

Further quantities used to characterize the system's structure are deduced from a cluster analysis. Particles are members of a cluster (of the same species) if the distance to their neighbors is smaller than $1.1\sigma$. 
Given the set of clusters in the system, 
we calculate for each cluster the maximum cluster size in $y$-direction. Averaging over all clusters and time yields the average elongation of clusters $\left<l_y\right>$ along the drive. In addition,
we check whether there exists (at least) one cluster which spans the system and is connected to its own mirror images, either in $x$- or $y$-direction. Given there exists such a cluster, we call the system 'percolated' in
$x$- or $y$-direction.

%-----DDFT-----
\subsection{\label{subsec:DDFT} Effective interactions in the driven system}
\begin{figure}
\includegraphics[width=0.8\columnwidth,left]{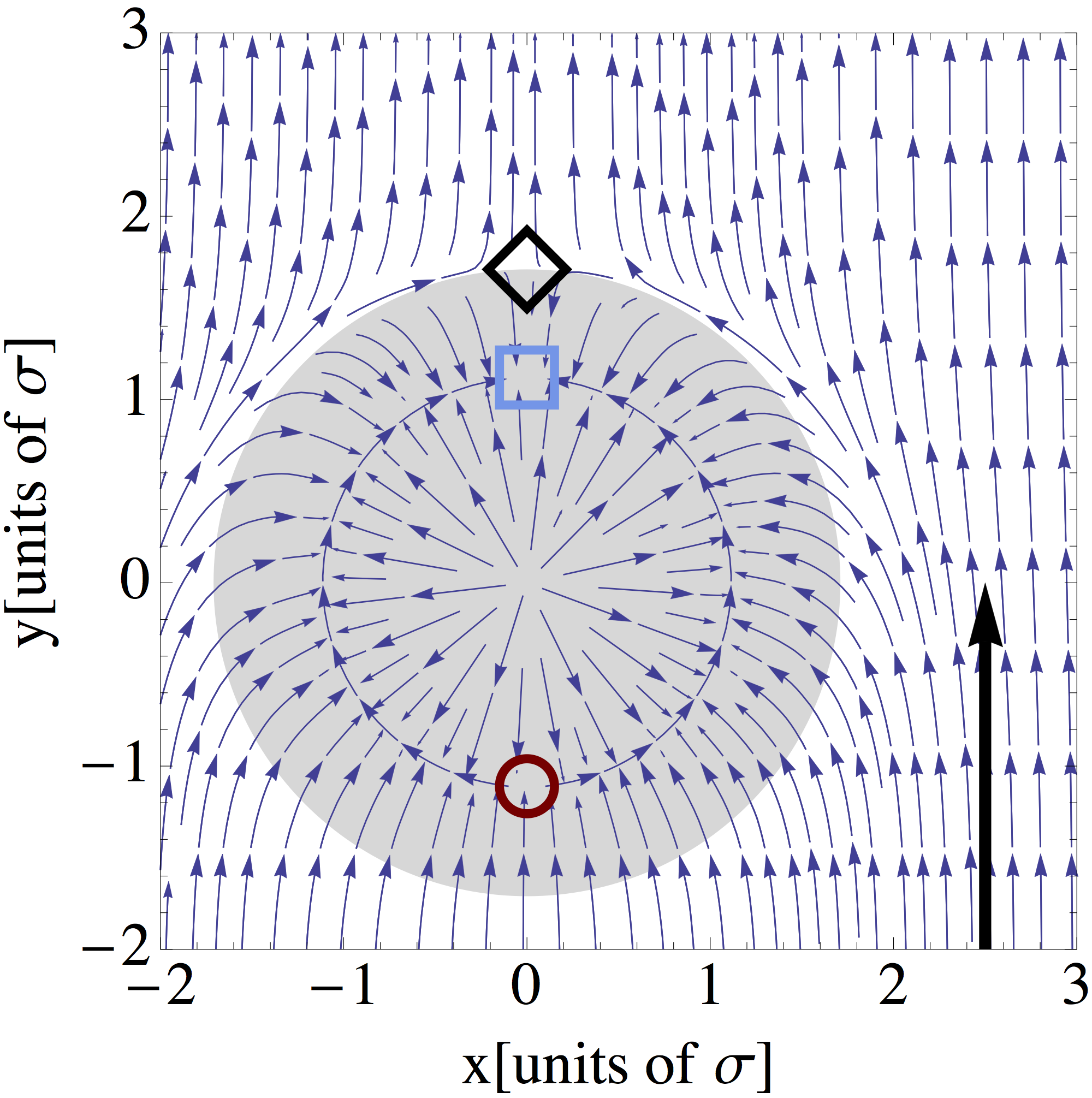}
\begin{picture}(0,0)
\put(20,+110){{\setlength{\fboxsep}{0pt}\colorbox{white}{\color{gray}\fbox{\includegraphics[width=0.4\columnwidth]{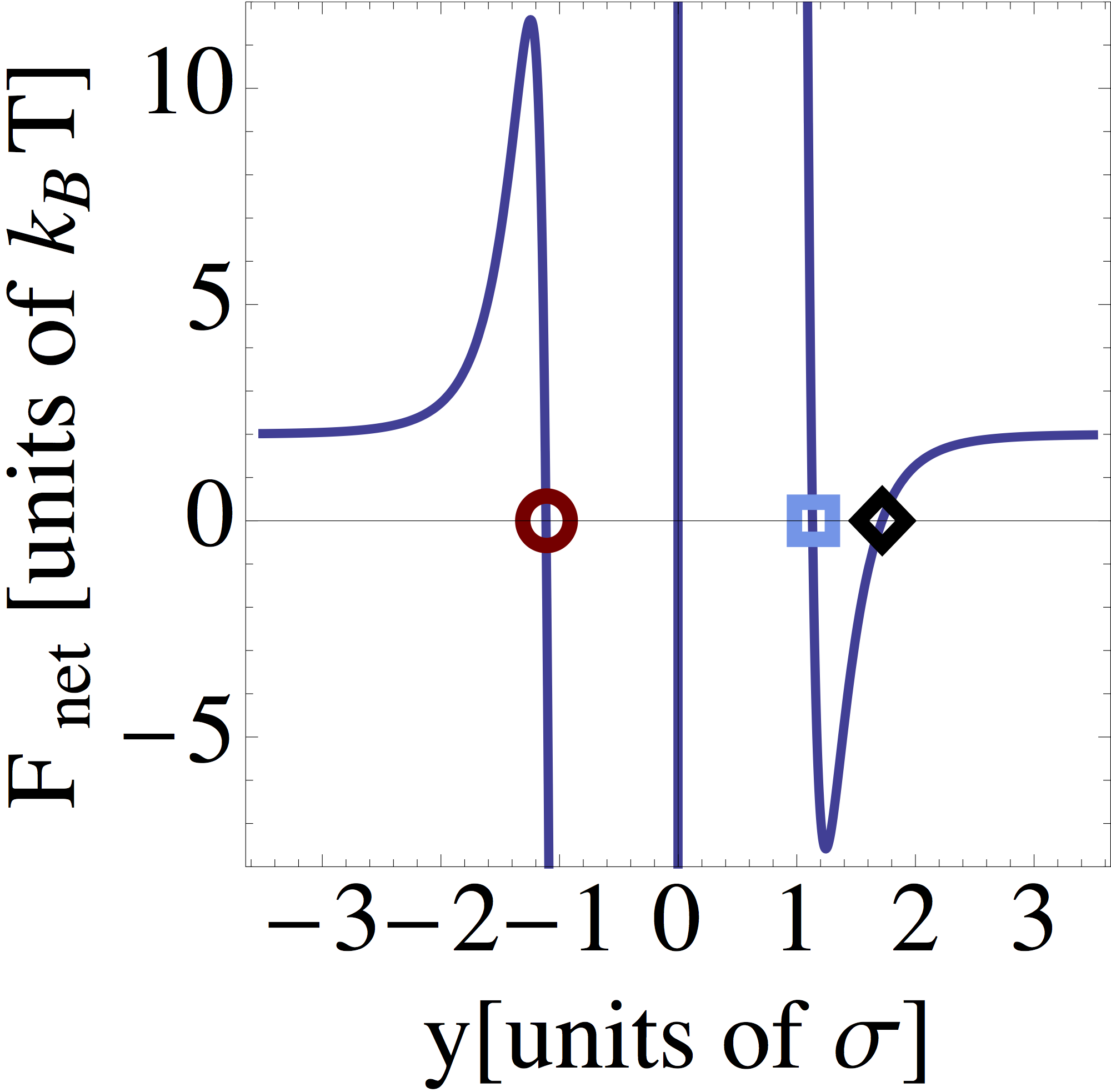}}}}}
\end{picture}
\caption{\label{fig:vectorField} Representation of the force field $\mathbf{F}^i_{\text{net}}$ at $\varepsilon^\ast=4$ and $f^\ast_\text{d}=2$ [see Eq.~(\ref{eq:forcebalance})] 
acting on a test particle $i$ with orientation $s_i=1$ due to the presence of a reference particle $j$ with orientation $s_j=-1$ in the center of the coordinate frame. 
Red circle, blue rectangle and black diamond indicate metastable, stable and unstable solutions
of $|\mathbf{F}^i_{\text{net}}|=0$, respectively. The test particle will approach the particle of opposite orientation from the bottom due to the driving force $f_d^\ast$ 
(indicated by the black arrow) and eventually reach the metastable configuration (red circle). This metastable arrangement will transform into the stable state (blue rectangle), 
a process which is driven by fluctuations. Once the particles are in the stable configuration, they can become separated by overcoming the effective 
potential $U_{\text{eff}}$ (see Eq.~(\ref{eq:effectivepotential}) and Fig.~\ref{fig:effectivePotential}). The inset shows the 
force acting on the particle as function of the $y$ position for $x=0$ relative to the fixed particle. Due to steric interactions, the area of $y\in[-1,1]$ is forbidden.}
\end{figure}
 
For two particles of different species, the attractive Lennard-Jones interaction competes with the propulsion force which drives these particles away from each other.
The net force experienced by a particle $i$ of species $s_i$ due to the presence of another particle $j$ of species $s_j$ is therefore: 
\begin{equation}
\label{eq:forcebalance}
\mathbf{F}^i_{\text{net}}(s_i,s_j,\mathbf{r}_{ij})=-\bm{\nabla} U_{\text{LJ}}(\mathbf{r}_{ij}) + \frac{1}{2}|s_i-s_j|\mathbf{f}_\text{d}(s_i) ~.
\end{equation}
In Fig.~\ref{fig:vectorField}, the net force $\mathbf{F}^i_{\text{net}}$ is shown as a force field experienced by a test particle $i$ (driven from bottom to top as indicated by the black arrow) 
due to the presence of a particle $j$ with fixed position in the center of the 
coordinate frame. Parameters are set to $\varepsilon^\ast=\varepsilon/k_BT=4.0$ and $f_d^\ast=f_d\sigma/k_BT=2.0$, and we consider the case $s_i\neq s_j$.
The inset of Fig.~\ref{fig:vectorField} shows the force field $\mathbf{F}^i_{\text{net}}$ along the driving force under the condition that both 
particles have equal x-coordinate (specifically, $x_i=x_j=0$).
Interestingly, there exist three solutions to $|\mathbf{F}^i_{\text{net}}|=0$, indicated in Fig.~\ref{fig:vectorField} by a red circle, a blue rectangle and a black diamond, corresponding to 
a metastable, a stable and an unstable solution, respectively. In the metastable configuration the $x$- and $y$-components of the force field are zero, but become finite even for very small 
displacements. In $x$-direction, the net force points away from the red circle, indicating an instability, while the $y$-direction is stable. 
The stable configuration is characterized such that the surrounding 
$x$- and $y$-components of the force field point towards the blue rectangle. Finally, in the proximity of the unstable configuration, the force field points always away from the black diamond.

Considering now a situation where a particle $i$ approaches particle $j$ from bottom to top (following the force field). First, the particle will reach the region around the metastable configuration (red circle) 
because of the general geometric setup. Then, it will be driven towards the stable configuration (blue rectangle) along the circular line. The latter
separates repulsive and attractive regions and connects, at the same time,
the metastable and stable configurations (see Fig.~\ref{fig:vectorField}, the inner circle with poles in red (circle) and blue (rectangle)).

\begin{figure}
\begin{center}
\includegraphics[width=0.8\columnwidth]{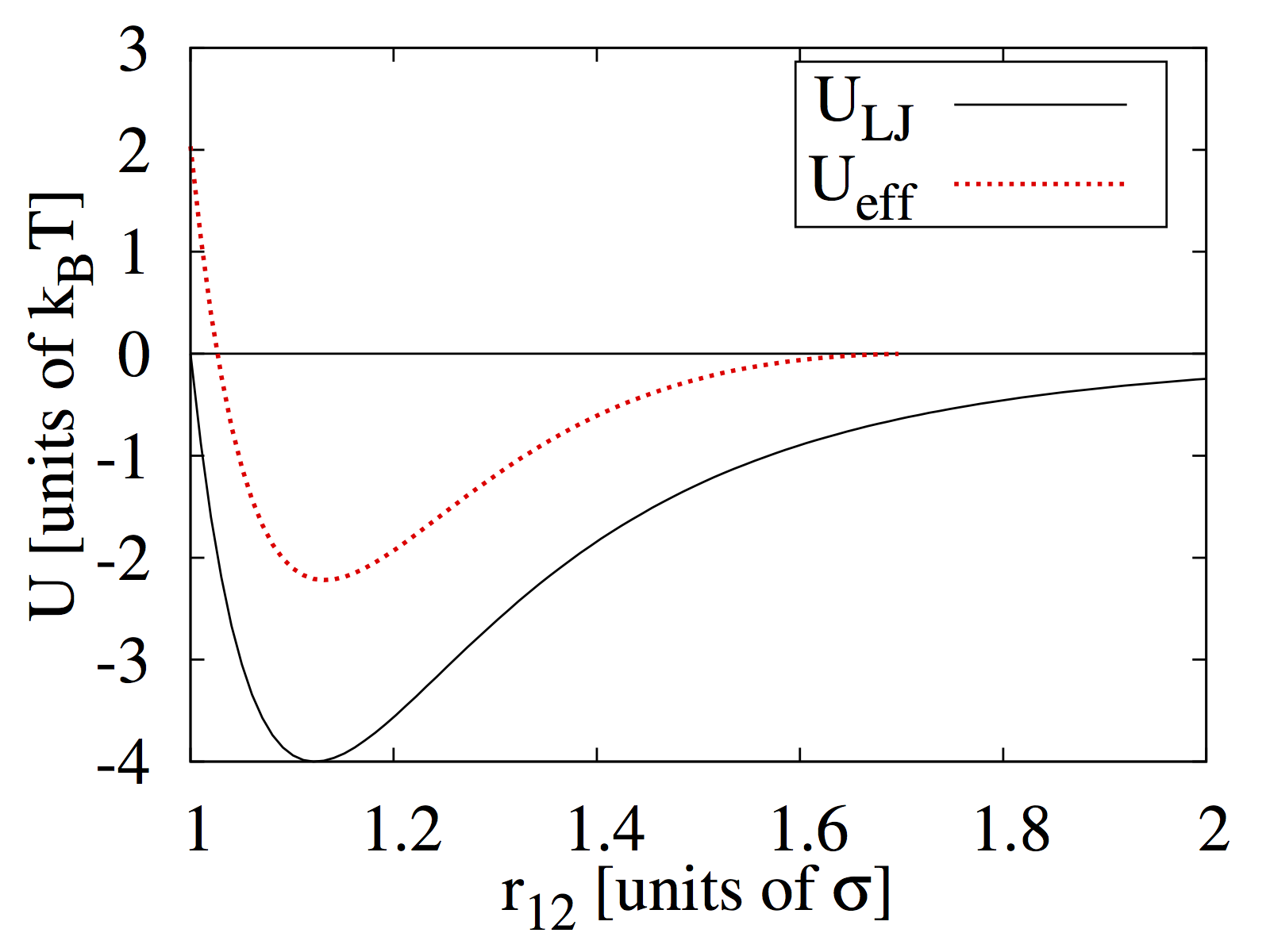}
\end{center}
\caption{\label{fig:effectivePotential} Effective potential $U_{\text{eff}}$ between particles of different species [see Eq.~(\ref{eq:effectivepotential})] shown as red dotted line for $f_d^\ast=2$ together with the LJ-interaction 
(black line) at $\varepsilon^\ast = 4$. In the presence of external driving, the strength and the range of attraction are decreased.}
\end{figure}

Once particle $i$ is in the stable configuration (blue rectangle), it can become separated from particle $j$ by overcoming the effective potential 
\begin{equation}
\label{eq:effectivepotential}
U_{\text{eff}}(r_{ij},s_i,s_j) = U_{\text{LJ}}(r_{ij}) - \frac{1}{2}|s_i-s_j|\left(f_\text{d}r_{ij} + U_0\right). 
\end{equation}
This effective interaction potential results from formal integration of Eq.~(\ref{eq:forcebalance}) 
along the $y$-axis under the conditions that $y_{ij}>0$ for $s_i=1$ ($y_{ij}<0$ for $s_i=-1$),  and $x_{ij}=0$.
The underlying assumption is that the two particles will always reach the stable configuration. 
Therefore the effective potential only depends on the distance in $y$-direction.
The constant $U_0$ is chosen such that $U_{\text{eff}}(r^\ast_{ij})=0$, where $r^\ast_{ij}$ corresponds to the locus of the black diamond in Fig.~\ref{fig:vectorField}, that is the 
distance at which the particles become free (in this one-dimensional picture $r^\ast_{ij}$ is the solution of $|\mathbf{F}^i_{\text{net}}|=0$).
Thus, the effective potential 
ranges from $r_{ij}=0$ to $r^\ast_{ij}$. %(from the middle of the fixed particle throught the blue circle, which is the stable configuration in contact, to the black circle, freedom point) 
For larger distances, at which particles are always driven apart, we set $U_{\text{eff}}(r_{ij})=0$ (see Fig.~\ref{fig:vectorField}). 
As an illustration, $U_{\text{eff}}(r_{ij})$ is plotted in Fig.~\ref{fig:effectivePotential} together with the LJ-potential. It is seen that the presence of the 
driving force decreases both, the strength and the range of attraction in our two-particle picture. A similar ansatz for an effective potential between swimming bacteria was recently 
introduced by Schwarz-Linek et al.~\cite{Schwarz-Linek}. Note that between particles of the {\it same} species, the {\it relative} driving force becomes zero. 
In this case, $U_{\text{eff}}$ reduces to the LJ-Potential (see Eq.~(\ref{eq:effectivepotential}) with $s_i=s_j$).

\subsection{\label{subsec:MethodsDDFT} Linear stability analysis}
In this study we supplement our simulations of the driven LJ system by a linear stability analysis in the framework of Dynamical Density Functional 
Theory~\cite{DDFTbasics}. Within this theory, the dynamics of the density fields $\varrho_\alpha(\mathbf r,t)$ are given by
\begin{equation}\label{eqn:DDFT}
\frac{\partial \varrho_\alpha (\mathbf r,t)}{\partial t} = (-\nabla) \cdot \left(-\frac{D}{k_\text{B}T}\varrho_\alpha(\mathbf r,t) \nabla \frac{\delta \mathcal F [\varrho_+,\varrho_-]}{\delta \varrho_\alpha(\mathbf r,t)}\right),
\end{equation}
where $\alpha$ is either $"+"$ or $"-"$ for the two different species. Further, $D$ is the diffusion constant and $\delta \mathcal F [\varrho_+,\varrho_-]/\delta \varrho_\alpha$ is the functional derivative of 
the Helmholtz free energy functional $\mathcal F$ with respect to the one-body density $\varrho_\alpha$. 
Equation~(\ref{eqn:DDFT}) refers to an overdamped system and involves an adiabatic approximation~\cite{adiabaticDDFT}. Here, we use the DDFT as a starting point for the stability analysis of the 'equilibrium' system
defined by the effective interaction potential $U_{\text{eff}}(r_{ij})$. Due to the attractive terms in $U_{\text{eff}}(r_{ij})$ one 
indeed expects instabilities related to fluctuations of the total density (e.g., condensation) or concentration (e.g., demixing)~\cite{lichtnerratchet}.

To this end, we consider small harmonic perturbations. 
We assume that the growth rate $\gamma(k)$ is the same for both species (see, e.g.~\cite{lichtnerbinarymagnetic}):
\begin{align}
\varrho_+(\mathbf r,t) &= \varrho_+^0 + \Delta \varrho(\mathbf r,t) = \varrho_+^0 + \phi e^{i\mathbf k\mathbf r}e^{\gamma(k) t} ~,\label{eqn:perturbations1}\\
\varrho_-(\mathbf r,t) &= \varrho_-^0 + \psi\Delta \varrho(\mathbf r,t) = \varrho_-^0+\psi\phi e^{i\mathbf k\mathbf r}e^{\gamma(k) t} ~.\label{eqn:perturbations2}
\end{align}
Here, $\Delta \varrho$ is a small density perturbation of amplitude $\phi$ and with wave number $|\mathbf k| = k$. 
The ratio for the perturbation amplitudes between the two species is denoted by $\psi$. 
A Taylor expansion of the free energy derivative up to linear order yields

\begin{align}\label{eqn:Expansion}
\begin{split}
&\frac{\delta \mathcal F[\varrho_+,\varrho_-]}{\delta \varrho_\alpha} = \left.\frac{\delta \mathcal F[\varrho_+,\varrho_-]}{\delta\varrho_\alpha}\right|_{\varrho_+^0,\varrho_-^0} \\
&+ \int d\mathbf r' \left.\frac{\delta^2 \mathcal F[\varrho_+,\varrho_-]}{\delta \varrho_\alpha\delta\varrho_\alpha}\right|_{\varrho_+^0,\varrho_-^0}(1-\delta_{\alpha,-})\Delta\varrho(\mathbf r',t) \\
&+\psi\int d\mathbf r' \left[\left.\frac{\delta^2\mathcal F[\varrho_+,\varrho_-]}{\delta\varrho_\alpha \delta\varrho_\alpha}\right|_{\varrho_+^0,\varrho_-^0}\delta_{\alpha,-}\right.\\
&\qquad\qquad\quad  \left.+\left.\frac{\delta^2\mathcal F[\varrho_+,\varrho_-]}{\delta\varrho_\alpha\delta\varrho_\beta}\right|_{\varrho_+^0,\varrho_-^0}(1-\delta_{\alpha,-})\right]\Delta\varrho(\mathbf r,t)\\
&+\int d\mathbf r'\left. \frac{\delta^2\mathcal F[\varrho_+,\varrho_-]}{\delta\varrho_\alpha\delta\varrho_\beta}\right|_{\varrho_+^0,\varrho_-^0}\delta_{\alpha,-}\Delta\varrho(\mathbf r',t)~. \quad (\beta\neq\alpha)
\end{split}
\end{align}

Here, $\delta_{\alpha,-}$ is the Kronecker delta, which is $1$ if $\alpha$ is $"-"$ and $0$ otherwise.

Inserting the expansion into Eqs.~(\ref{eqn:DDFT}) results into two coupled differential equations (see Eqs.~(\ref{DGL-DDFT1}),~(\ref{DGL-DDFT2}) in the Appendix). Importantly, only the
second functional derivative of the free energy functional appears in these equations. The latter relates to the direct correlation functions $c_{\alpha\beta}^{(2)}(|\mathbf r-\mathbf r'|;\varrho_+^0,\varrho_-^0)$ 
of the unperturbed system via
\begin{equation}
\label{DCF-free}
k_\text{B}Tc_{\alpha\beta}^{(2)}(|\mathbf r-\mathbf r'|;\varrho_+^0,\varrho_-^0) = -\frac{\delta^2\mathcal F_{ex}[\varrho_+,\varrho_-]}{\delta\varrho_\alpha(\mathbf r')\delta\varrho_\beta(\mathbf r)}~.
\end{equation} 
Here we employ the random phase approximation~\cite{Hansen}, 
\begin{equation}
\label{randomphaseDCF}
c_{s_i,s_j}(r) \approx c_0(r) -\beta w(r,s_i,s_j)~,
\end{equation}
where $c_0(r)$ is the direct correlation function of hard
spheres with diameter $\bar\sigma$, defined as the distance below which $U_{\text{eff}}$ becomes positive. This function is known analytically~\cite{HardSphereCorrelationFunction}. 
Further, $w(r,s_i,s_j)$ corresponds to the effective potential [see  Eq.~(\ref{eq:effectivepotential})] and $\beta = (k_\text{B}T)^{-1}$. 
Using a Fourier transform with respect to the particle coordinates and assuming that the transformed correlation functions $\tilde c^{(2)}_{\alpha \beta}$
and the function $\gamma(k)$ only depend on the magnitude of $k$ (which is reasonable since we are expanding around a homogeneous state) we obtain two equations in momentum space:
\begin{align}
\begin{split}
\gamma(k) \Delta \varrho(\mathbf k,t) &= -k^2\Gamma \Delta \varrho(\mathbf k,t) [1-\varrho_+^0\tilde c_{++}^{(2)}(k;\varrho_+^0,\varrho_-^0)\\
&-\psi\varrho_+^0c_{+-}^{(2)}(k;\varrho_+^0,\varrho_-^0)]~,
\end{split}\label{eqn:momentumSpace1}
\\
\begin{split}
\psi\gamma(k)\Delta\varrho(\mathbf k,t) &=-k^2\Gamma\Delta\varrho(\mathbf k,t)[\psi-\psi\varrho_-^0\tilde c_{--}^{(2)}(k;\varrho_+^0,\varrho_-^0)\\
&+\varrho_-^0\tilde c_{-+}^{(2)}(k;\varrho_+^0,\varrho_-^0)]~, 
\end{split}\label{eqn:momentumSpace2}
\end{align}
where $\Gamma = D/k_\text{B}T$. The two equations are fulfilled simultaneously, if
\begin{align}\label{eqn:gamma}
\begin{split}
\gamma(k) =&  \frac{k^2\Gamma}{2}\left(\tilde c_{--}^{(2)}\varrho_-^0+\tilde c_{++}^{(2)}\varrho_+^0 -2\right)\\
&\pm \frac{k^2\Gamma}{2}\left[\left(\tilde c_{++}^{(2)}\varrho_+^0\right)^2+\left(\tilde c_{--}^{(2)}\varrho_-^0\right)^2\right.\\ 
&\left.+4\tilde c_{+-}^{(2)}\tilde c_{-+}^{(2)}\varrho_+^0\varrho_-^0 -2\tilde c_{--}^{(2)}\tilde c_{++}^{(2)}\varrho_+^0\varrho_-^0\right]^{1/2}
\end{split}
\end{align}
as shown in the Appendix [see Eq.~(\ref{eqn:Matrix}) and~(\ref{eqn:gamma_appendix})]. 
In Eq.~(\ref{eqn:gamma}), we have dropped the arguments of the Fourier transforms of the direct correlation functions $\tilde c^{(2)}_{\alpha \beta}$.

For positive values of $\gamma (k)$, Eqs.~(\ref{eqn:perturbations1}) and~(\ref{eqn:perturbations2}) reveal that the 
density perturbations $\Delta \varrho$ with wave number $k$ grow exponentially in time. 
Equation~(\ref{eqn:gamma}) can be simplified by noting that the two species are identical and thus, $\tilde c_{+-}^{(2)}=\tilde c_{-+}^{(2)}$ and $\tilde c_{++}^{(2)}=\tilde c_{--}^{(2)}$. 
Also, since we are considering a $50:50$ binary mixture, $\varrho_+^0=\varrho_-^0=\varrho/2$. Thus Eq.~(\ref{eqn:gamma}) reduces to
\begin{equation}\label{eqn:Gamma}
\gamma(k)_{\pm} = -k^2\Gamma\left(1-\frac{\varrho}{2} \tilde c_{++}^{(2)}\right)\pm k^2\Gamma \tilde c_{+-}^{(2)}\frac{\varrho}{2}~.
\end{equation}

We note that in the "long-wavelengths" limit $k\to 0$, the quantities $A_{\pm}=-\gamma_{\pm}(k)/\Gamma k^2$ become identical to the quantities appearing in the Kirkwood-Buff theory \cite{KirkwoodBuff1951} for a system's spinodal. Stability then means that $A_{\pm}(k=0)$ is positive.

%%******Results*****%%
\section{\label{sec:Results}Results}
\subsection{Overview of BD simulation results\label{subsec:OverviewSimulation}}
\begin{figure}[]
\includegraphics[width=\columnwidth]{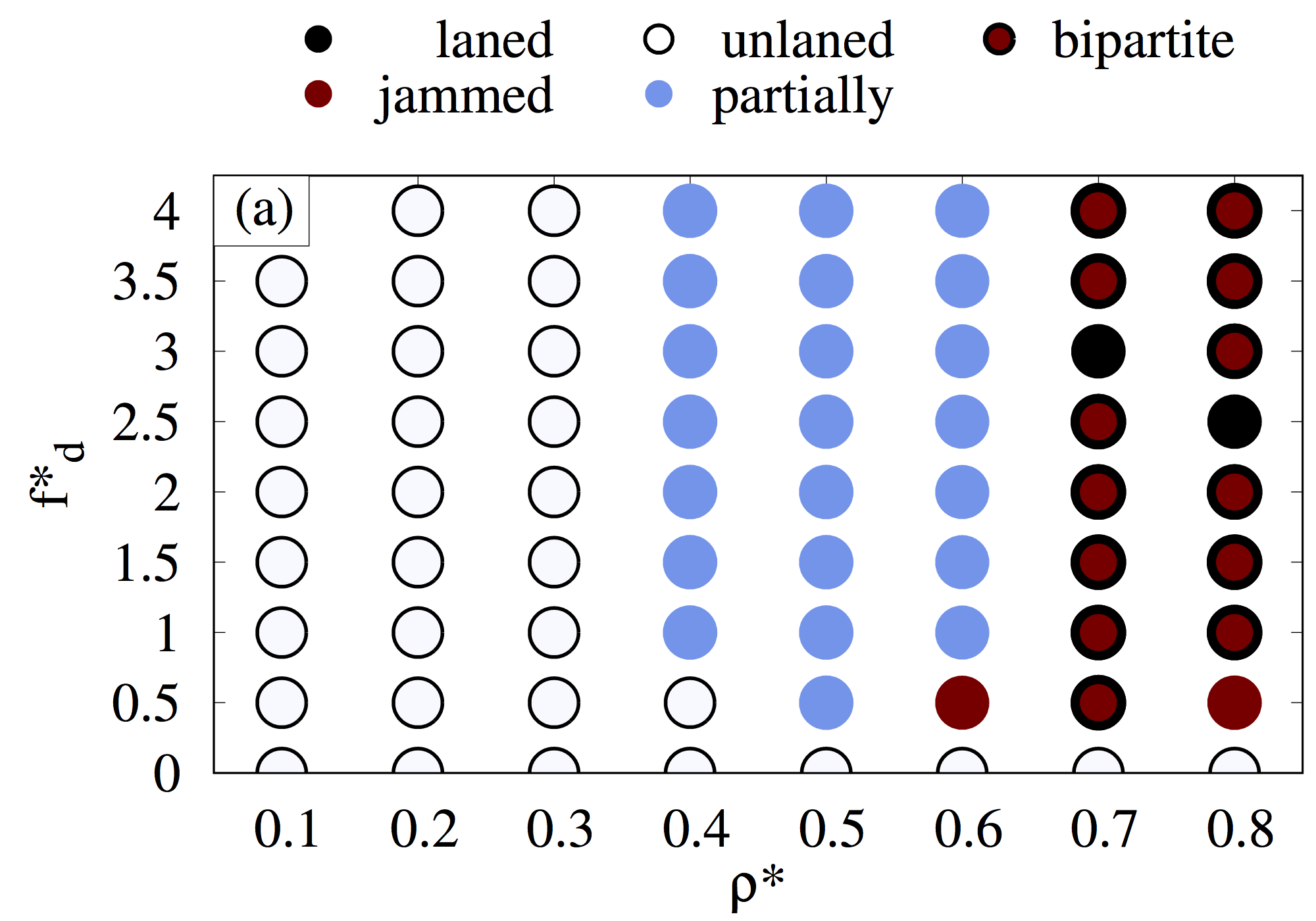}\\
\includegraphics[width=\columnwidth]{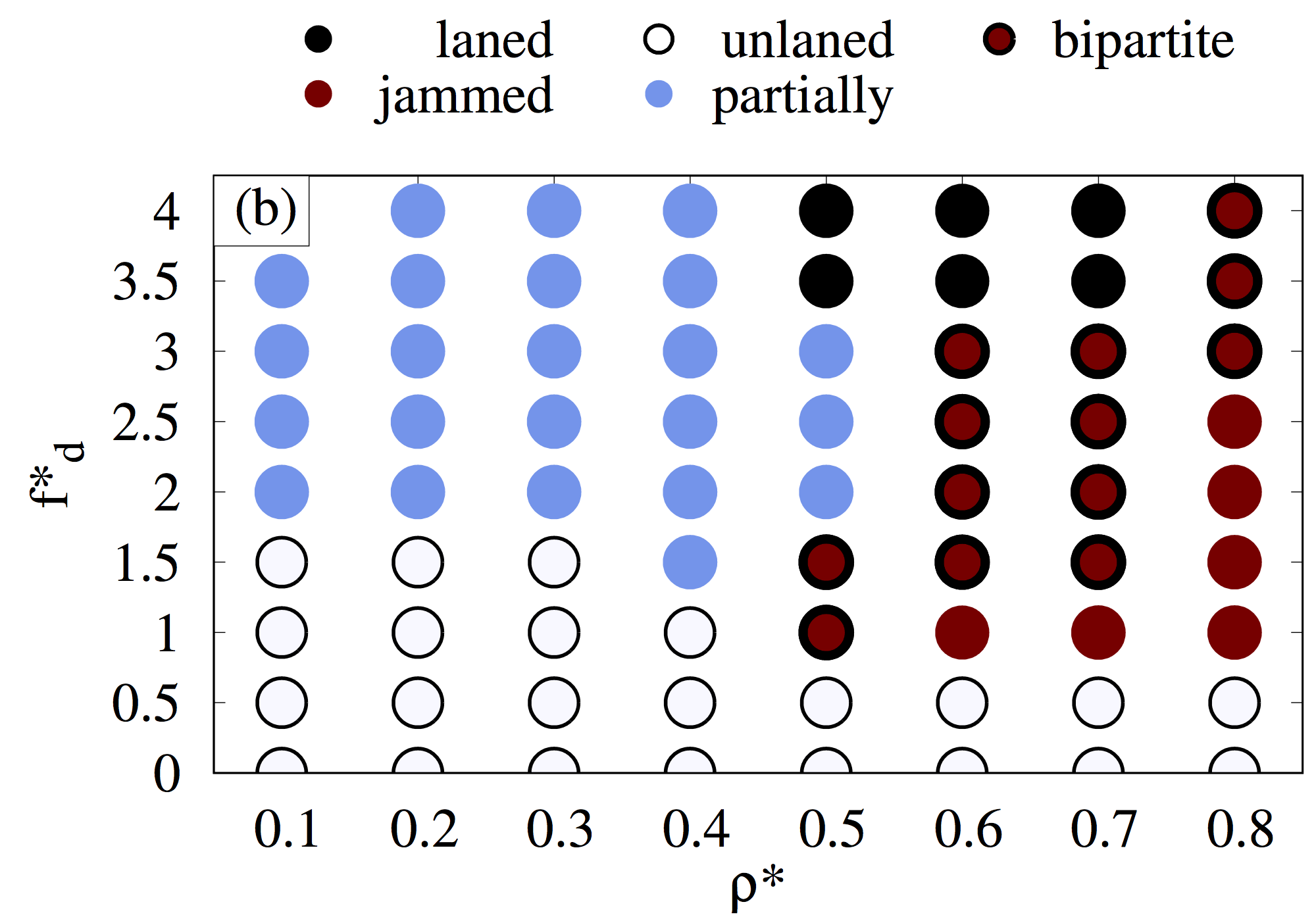}
\caption{\label{fig:StateDiagramm1} State diagram from BD simulations at (a) $\varepsilon^\ast = 2.5$ and (b) $\varepsilon^\ast =4$ 
in the $\varrho^\ast$-$f^\ast_\text{d}$-plane. The data points are colored according to the four different states as 
discussed in the text: laned (I, black), jammed (II, red (dark gray)), unlaned (III, white) and partially laned (IV, blue (light gray)). Furthermore we indicated bipartite simulation runs 
by red (dark gray) and black color code (see Sec. \ref{subsec:StructureAnalysis4}). Representative snapshots of the different states are shown in Fig.~\ref{fig:Snapshots}.}
\end{figure}
\begin{figure}
\includegraphics[width=\columnwidth]{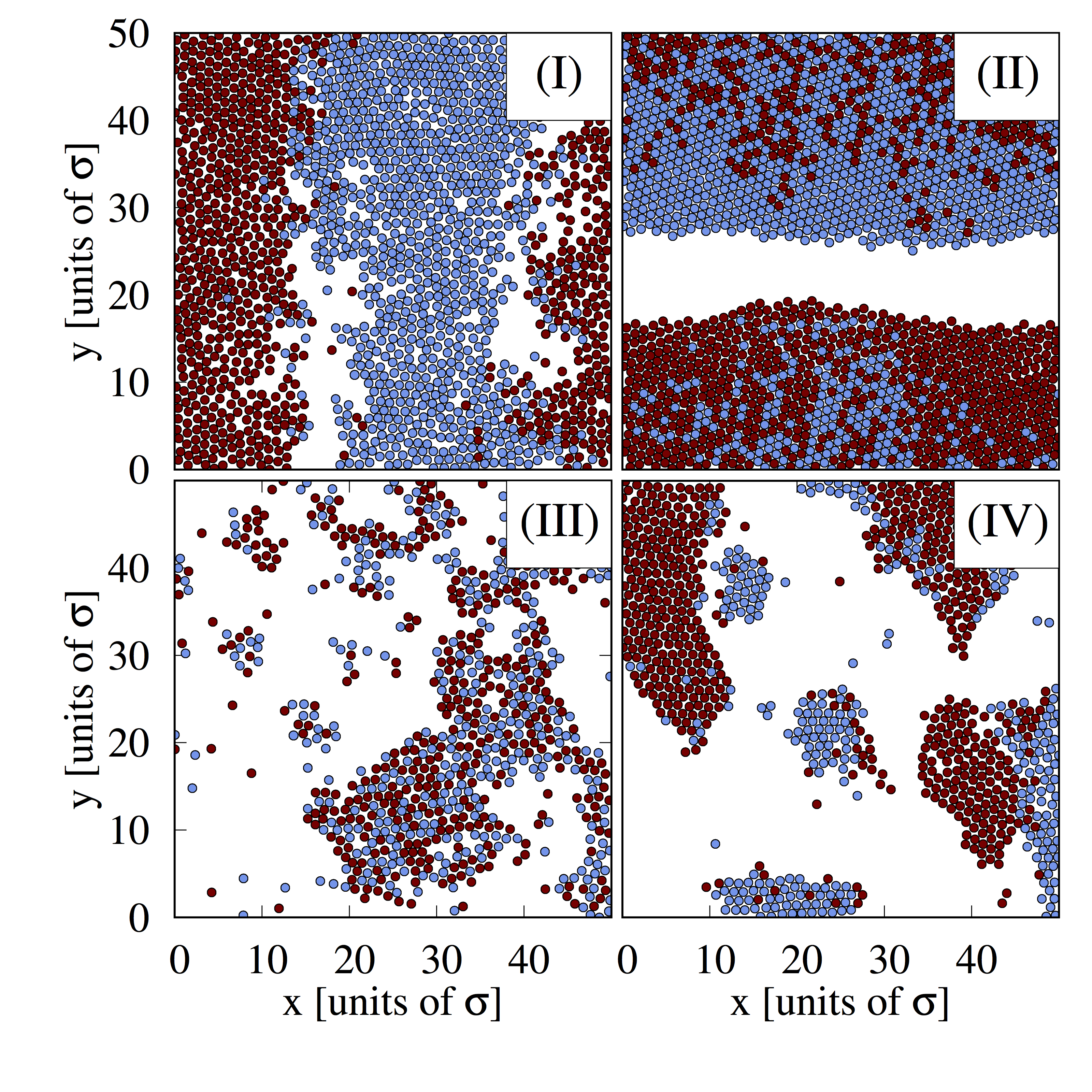}
\caption{\label{fig:Snapshots} Snapshots obtained from BD simulations illustrating the four different states found in the driven system: 
(I) laned state ($\varrho^\ast = 0.7,~\varepsilon^\ast=2.5,~f^\ast_\text{d} = 4.0$), 
(II) jammed state ($\varrho^\ast = 0.8,~\varepsilon^\ast=4.0,~f^\ast_\text{d} = 3.5$), 
(III) non-laned state ($\varrho^\ast = 0.3,~\varepsilon^\ast=2.5,~f^\ast_\text{d} = 0.5$) and 
(IV) partially laned state ($\varrho^\ast = 0.4,~\varepsilon^\ast=4.0,~f^\ast_\text{d} = 4.0$). 
The parameter sets considered here correspond to data points in Fig.~\ref{fig:PhiVsY} encircled in grey.} 
\end{figure}
We performed BD simulations for a large range of reduced densities $\varrho^\ast=\rho\sigma^2$ and driving forces $f_d^\ast$. In Fig.~\ref{fig:StateDiagramm1} we plot corresponding state 
diagrams for the two interaction strengths $\varepsilon^\ast = 2.5$ and $\varepsilon^\ast = 4$, respectively.  Both values of $\varepsilon^\ast$ correspond to the sub-critical regime of the undriven LJ system, 
which has a critical point at $\varepsilon_c^\ast\approx2.17$ and $\rho^\ast_c\approx0.35$~\cite{smit_frenkel:LJ2D}. 
Further, these values have previously been shown to yield  significantly enhanced lane formation as compared to systems with solely repulsive pair interactions~\cite{kogler1}.
In Fig.~\ref{fig:StateDiagramm1}, colored symbols indicate different types of laned states described in detail in section \ref{subsec:StructuralProperties}. 
The main criteria for identifying the different states are the laning order parameter $\phi$ and the mean extension $\left<l_y\right>$ of clusters along the direction of the driving. 
Generally, we find that at fixed driving force, an increase of the density and/or the interaction strength enhances the systems ability to form lanes.
For the special case $\rho^\ast=0.5$ the present results are consistent with those reported in~\cite{kogler1}.

The different types of laned states are illustrated in
Fig.~\ref{fig:Snapshots} by BD snapshots. There exist perfectly laned (I), jammed (II), non-laned (III) and an partially laned states (IV).
In short, the partially laned states are characterized by relatively large particle aggregates of mostly one particle species, which are not percolated along 
the driving force and undergo collisions frequently. The overall structure, including the number and position of lanes, changes rapidly. 
In contrast to that, we find that the perfectly laned state (I) is very stable over
time with the lanes being percolated along the driving force.
The jammed state (III) is also percolated, but perpendicular to the drive. Please note, that the term 'percolation' is used if a cluster of particles of 
the same species spans the entire system. Particles of opposite species might be trapped inside such a cluster, but percolation requires
an unbroken sequence of neighboring particles of the same type.
Finally, the non-laned state in Fig.~\ref{fig:Snapshots} (III) is characterized by aggregates composed of both particle species.

As seen from Fig.~\ref{fig:StateDiagramm1}, for a small interaction strength ($\varepsilon^\ast=2.5$) laning starts at intermediate densities ($\varrho^\ast\gtrsim 0.3$).
However,  perfectly laned states only occur for very high densities ($\varrho^\ast\approx0.7$). 
For an interaction strength of $\varepsilon^\ast=4$, laned states are found for all densities considered here, but perfect lanes
again occur only for higher densities ($\varrho^\ast\gtrsim0.5$).
A common feature visible at both values of $\varepsilon^\ast$ is that lanes appear already at very small driving forces compared to purely repulsive systems~\cite{finite_size:lane, loewen:DFT-phasediagram}.
Starting from this general overview, we now proceed with a more quantitative discussion.

\begin{figure}[] 
\includegraphics[width=\columnwidth]{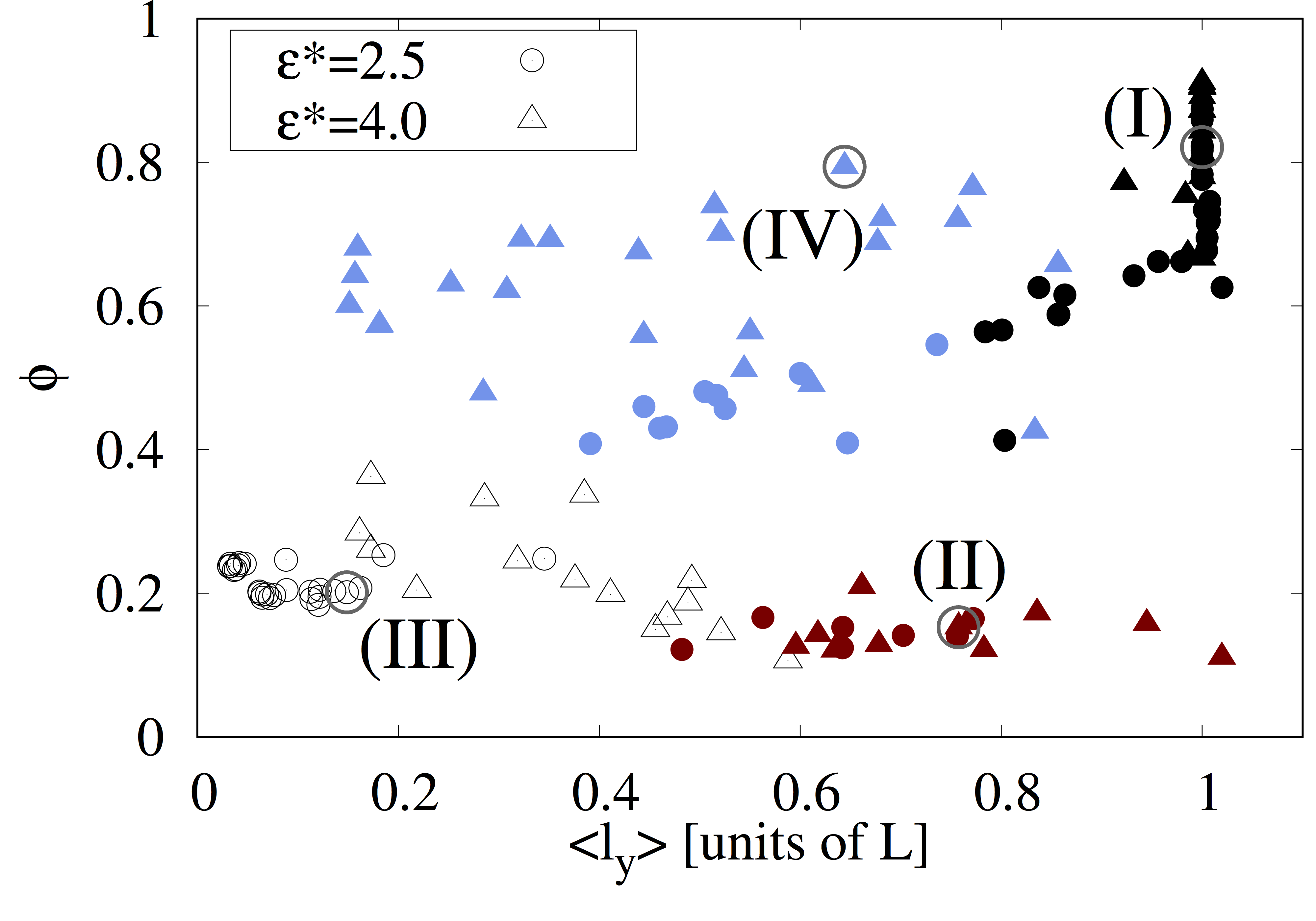}
\caption{\label{fig:PhiVsY} Laning order parameter $\phi$ against the mean elongation of 
clusters in y-direction, $\left<l_y\right>$. Results from BD simulations at various densities 
$\varrho^\ast$ (from  $0.1$ to $0.8$) and driving forces $f^\ast_\text{d}$ (from $0.5$ to $4.0$) 
categorized into the following states: (I, black) perfectly laned state, (II, red (dark gray)) jammed state, (III, white) 
non-laned state and (IV, blue (light gray)) partially laned state. Circles (triangles) correspond to 
the interaction strength $\varepsilon^\ast=2.5~(4.0)$. Encircled data points correspond to snapshots in Fig.~\ref{fig:Snapshots}.}
\end{figure}

In Fig.~\ref{fig:PhiVsY} we show the degree of laning $\phi$ against the quantity $\left< l_y\right>$.
This unusual representation is chosen here, because the laning order parameter $\phi$ 
alone turns out to be not sufficient to characterize the different types of lanes.  
Note that the data points in Fig.~\ref{fig:PhiVsY} correspond to simulation results at 
various combinations of the density (from $\varrho^\ast = 0.1$ to $\varrho^\ast = 0.8$), 
the driving force $f^\ast_\text{d}$ (from $0.5$ to $4.0$), and the interaction strengths ($\varepsilon^\ast=2.5$ and $4$).
The four states I-IV are indicated by different colors.
In the following sections~\ref{subsec:StructureAnalysis1}-\ref{subsec:StructureAnalysis4}, we discuss the structural properties of the different states 
based on order parameters and the pair correlation functions $g^s(x)$ and $g^s(y)$, see Eq.~(\ref{eqn:RadialDistributionX}) and Fig.~\ref{fig:RadialDistribution}.
These functions allow to characterize not only the local structures of aggregates, but also the properties of the entire lanes at large distances.

\subsection{\label{subsec:StructuralProperties} Structural properties}
\subsubsection{\label{subsec:StructureAnalysis1} Perfectly laned states}
\begin{figure}
\includegraphics[width=\columnwidth]{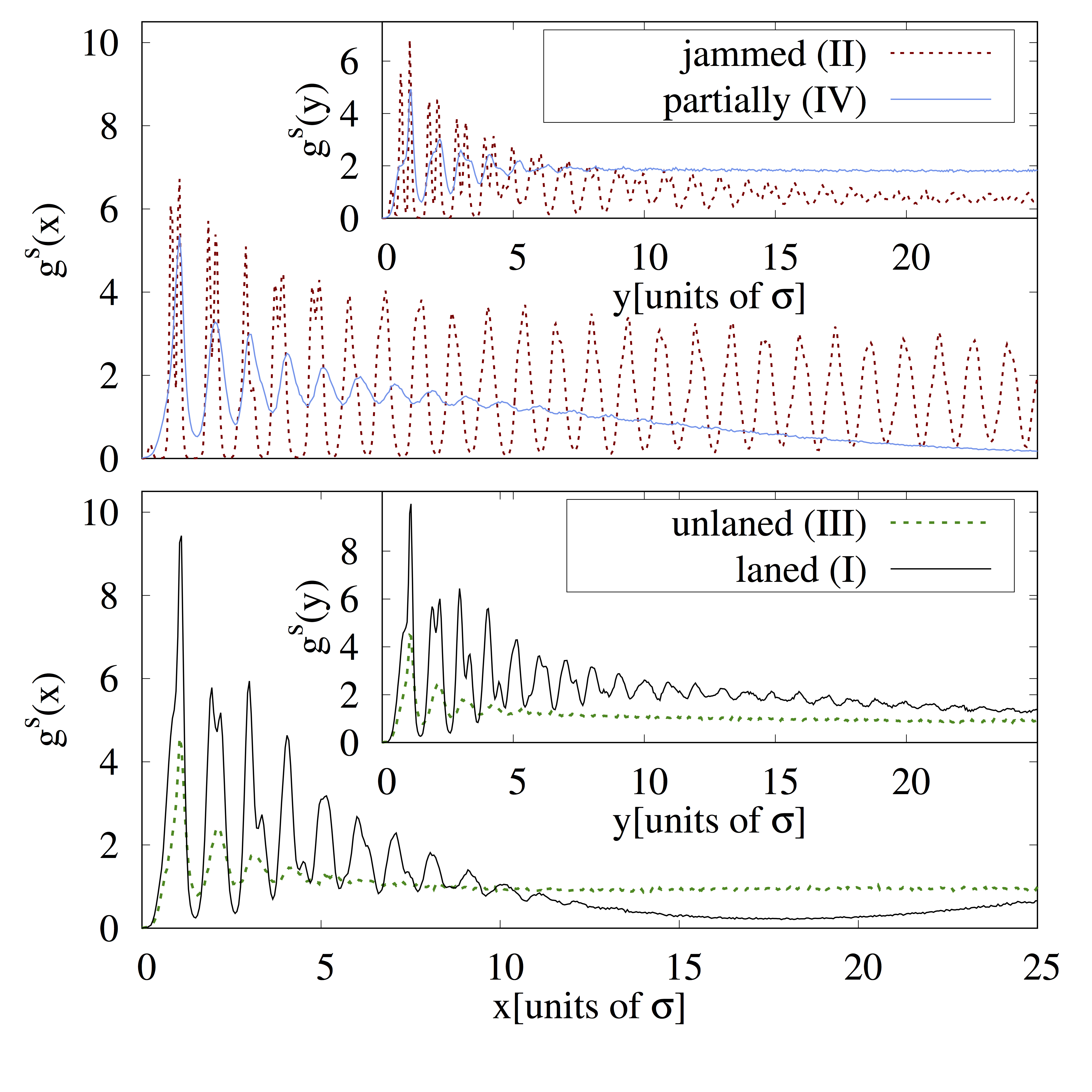}
\caption{\label{fig:RadialDistribution} Pair correlation functions (PCF) between particles of the same type characterizing 
the internal structure of the different states. The main top figure shows the PCF in $x$-direction for the jammed state (II, red dotted) and the partially laned state (IV, blue (light gray)), the inset the corresponding PCF in $y$-direction. The bottom figures display the PCF in $x$- and $y$-direction of 
the unlaned state (III, green dashed) the laned state (I, black) 
and in the main figure and the inset, respectively. 
The PCF are calculated from the simulation runs from which the snapshots in Fig.~\ref{fig:Snapshots} are taken.}
\end{figure}

%%%%%%%%%%%PERFECT LANES
The perfectly laned states (I) are characterized by large values of $\phi\gtrsim0.5$ and $\left<l_y\right>\gtrsim0.8$ (see black symbols in Fig.~\ref{fig:PhiVsY}). 
In this state, lanes are essentially large elongated clusters, which are composed of only one particle species. 
These elongated clusters are percolated in the direction of the driving force (i.e. the $y$-direction), 
yielding macroscopic system-spanning lanes [for a representative snapshot see Fig.~\ref{fig:Snapshots} (I)].
The characteristic features of perfectly laned  states are also reflected by the pair correlation functions between particles of the same type, see Fig.~\ref{fig:RadialDistribution}.
Specifically, the function $g^s(x)$ (black) displays a minimum at $x \gtrsim25 \sigma$. The latter can be interpreted as the center of the neighboring lane, which gives a lower limit for the width of lanes.
On smaller distances, the regular peaks of $g^s(x)$ reflect the pronounced local ordering of the particles suggesting a crystalline internal structure of lanes.
Considering the pair correlation function $g^s(y)$ shown in the inset of the top of Fig.~\ref{fig:RadialDistribution}, which is along the driving force and along the lanes, we find a different behavior. 
Here the function tends to the value $2$, indicating that the vast majority of particles found in $y$-direction is of the same species. 
However, there is no long ranged ordering inside the lanes, as $g^s(y)$ smoothly approaches its long ranged limit. 
Thus, the lanes can not be interpreted as a system-spanning {\it crystalline} structure. Rather they form an overall fluid-like structure consisting of ordered clusters.

In summary, the properties of the laned states in our system are consistent with
the classical picture of lane formation in driven hard-sphere (HS) systems~\cite{finite_size:lane}. 
However, in the present case the clusters are semi-crystalline aggregates in contrast to the fluid-like structures found in HS-systems
(although a special type of crystalline hexagonal ordering has also been observed in HS-systems~\cite{loewen_order:lane}).
We also note, that at high densities ($\varrho^\ast \gtrsim 0.6$) only two lanes emerge, 
one for each species. In this situation, the laning is associated to a complete demixing of  the two species. 

\subsubsection{\label{subsec:StructureAnalysis2} Unlaned states}
The unlaned states (III) (see transparent symbols in Fig.~\ref{fig:PhiVsY}) are characterized by 
small values of the laning order parameter $\phi \lesssim 0.4$, while the cluster sizes  can be relatively large (up to $\left<l_y\right> \approx 0.6 $), particularly at larger densities. 
Within the unlaned states, $g^s(x)$ and $g^s(y)$ display typical features of liquid and/or cluster phases
(see Fig.~\ref{fig:RadialDistribution}) in the sense that the correlations are of short range and decay rapidly in magnitude for $x\rightarrow \infty$. 
In this limit, both functions tend to $1$, indicating a homogeneous mixture of the two particle species. %\deleted{states are characterized by significantly smaller values of $g^s(x)$ and greater values of $g^s(y)$ in the long-ranged limit.}
Thus, the system is neither demixed nor crystallized, consistent with the simulation snapshot in Fig.~\ref{fig:Snapshots} (III).
 
\subsubsection{\label{subsec:StructureAnalysis3} Partially laned states}

In addition to the extreme cases of perfect laning (I) and no laning (III), we also find
partially laned states (IV), which are more difficult to characterize. 

As seen from Fig.~\ref{fig:PhiVsY} (blue (light gray) symbols) the system here still yields significant values of $\phi \geq 0.4$ but is not percolated along the driving force.
Further, the 'length of lanes' measured by $\left<l_y\right>$ varies strongly from $\left<l_y\right>\approx0.1$ to $\left<l_y\right>\approx0.9$.
This strong variation of $\left<l_y\right>$ reflects that the particles form (hexagonal) aggregates of various size (depending on the density and temperature), 
which are moving on 'empty' or 'free' lanes [see Fig.~\ref{fig:Snapshots} (IV)]. 
We name these states 'partially', because clusters of different particle species 
frequently collide, as they diffuse into other lanes, dissolve and reconfigure. Hence, it is often the case that only \textit {parts} of the system display laned structures.

Also regarding the local structure, the partially laned states appear as an intermediate case with structural properties from both, unlaned and laned states. 
On the one hand, the pair correlation perpendicular to the drive $g^s(x)$ (see bottom main Fig.~\ref{fig:RadialDistribution}), shows a minimum at $x \approx 17\sigma$, 
similar to the perfectly laned state. 
On the other hand, the function $g^s(y)$ is only slightly larger 
than $1$ for large distances, similar to the unlaned state. Thus, the system 
shows no system spanning lanes.
The 'laning-like' behavior of $g^s(x)$ results from the appearance of isolated clusters of different 
sizes (which strongly depend on the overall density) which move on 'empty' lanes and diffuse between neighboring lanes.
From the preceding discussion it is obvious, that the characterization of partially laned states is somewhat
arbitrary as the transition from unlaned to laned states
is very gradual. Nevertheless, it should be noted that the very appearance of inhomogeneously 
structured lanes seems to be a unique feature of attractive systems such as dipolar microswimmers~\cite{kogler1} or the LJ fluid considered here. 
In fact, the 'classical' driven HS systems display either homogeneous or completely crystalline structures within the lanes.   

\subsubsection{\label{subsec:StructureAnalysis4} Jammed states}

The jammed states (II) (see red (dark gray) symbols in Fig.~\ref{fig:PhiVsY}) are characterized by small values 
of the laning order parameter ($\phi\lesssim 0.3$) and large values of the elongation ($\left<l_y\right> \gtrsim0.5$). 
These states are percolated in $x$-direction, 
that is, perpendicular to the drive, and they occur solely at high densities ($\varrho^\ast \geq 0.5$). 
This suggests a relation to the freezing transition of the equilibrium Lennard-Jones fluid, which takes place at 
high densities and/or low temperatures (specifically the triple point density and temperature are given by $\varrho^\ast\approx0.75$ and $\varepsilon^\ast \approx 2.4$~\cite{Feng2000}). 
In the non-equilibrium, high-density situation considered here, particles moving in opposite 
directions can not pass each other without colliding with other particles, which are most probably particles of the other species. 
Due to the strongly attractive interactions this might prevent a demixing of particle species. Eventually, the 
system forms one large aggregate [see snapshot in Fig.~\ref{fig:Snapshots} (II)], which is composed of two half’s, each dominated by 
one particle species. Thus, the jammed state is only  partially demixed, with the (smeared) border 
between its parts being orientated perpendicular to the driving force. 
This is also reflected by the pair correlation functions in $x$- and $y$-direction 
shown in the top of Fig.~\ref{fig:RadialDistribution}. Specifically, $g^s(x)$ reflects crystalline ordering perpendicular to the drive throughout 
the whole system. Along the drive, the function $g^s(y)$ tends to $1$ for large distances, indicating that the particle species are not demixed.
Further, the double peaks of the correlation functions along both directions 
suggest an hexagonal-like ordering on the local scale, in accordance with the visual observation from Fig.~\ref{fig:Snapshots} (II). 
Overall, the jamming found here reminds of an arrested phase separation (demixing) 
as it is found, e.g., in attractive glasses~\cite{oneliquidtwoglasses}.

\begin{figure}[]
\includegraphics[width=\columnwidth]{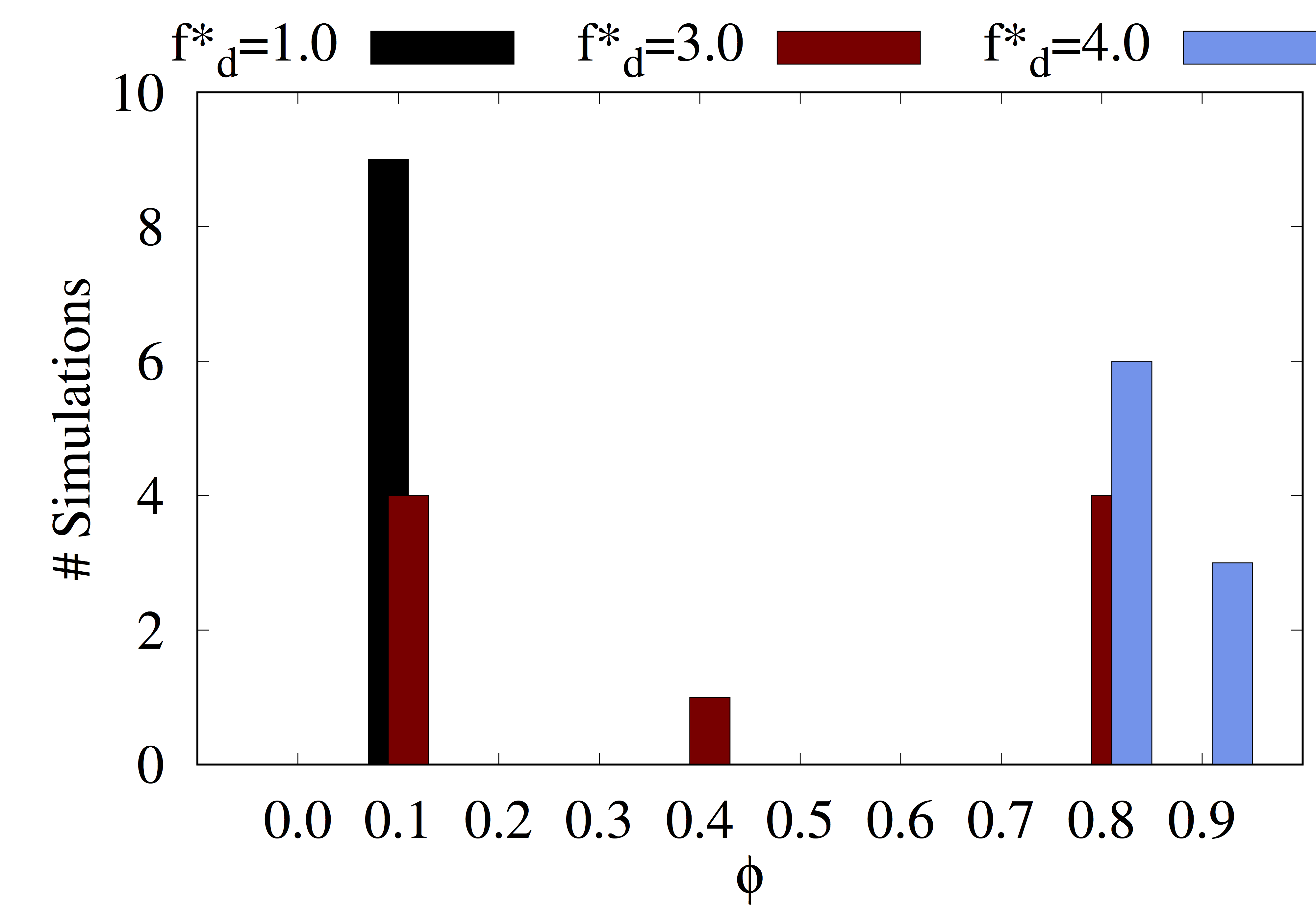}
\begin{picture}(0,0)
\put(-41,+75){\includegraphics[width=0.49\columnwidth]{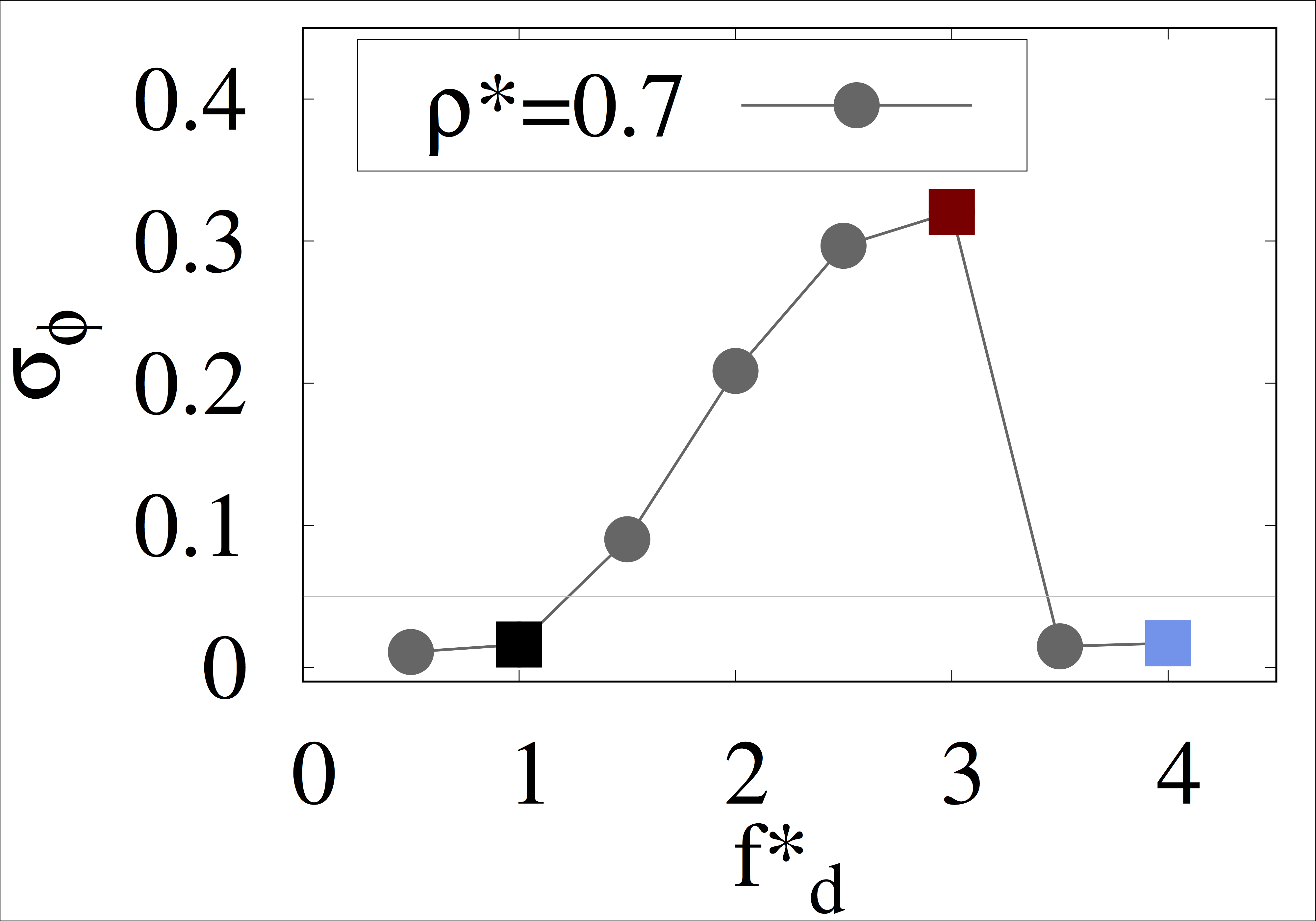}}
\end{picture}
\caption{\label{fig:Bipartite}Main figure: 
Distribution of the laning order parameter $\phi$ for different driving forces at fixed $\varepsilon^\ast=4.0$ and $\varrho^\ast=0.7$. 
For a small driving force ($f^\ast_\text{d}=1.0$, black bar) all simulation runs end up in a jammed state whereas for a large driving 
force ($f^\ast_\text{d}=4.0$, blue (light gray) bars) all simulation runs achieve a laned state. For medium driving forces ($f^\ast_\text{d}=4.0$, red (dark gray) bars) some simulation 
runs end up in a laned state and some in a jammed state. Inset: Standard deviation $\sigma_\phi$ of $\phi$ 
from different simulation runs. The system is called bipartite if $\sigma_\phi\geq 0.05$. Rectangular markers give the 
standard deviation of the distributions shown in the main figure in the same color code.}
\end{figure}

We should note that the occurrence of jamming in our system strongly depends on the value of the driving force, as well as on the initial conditions.
To illustrate the dependence on $f^\ast_{\text{d}}$, we provide in Fig.~\ref{fig:Bipartite} the distribution of $\phi$ at 
fixed interaction strength $\varepsilon^\ast = 4.0$ and density of $\varrho^\ast = 0.7$ for different driving forces. 
For small driving forces, e.g. $f^\ast_\text{d} = 1.0$, the system always runs into a jammed state, which is indicated by a small 
laning order parameter of $\phi = 0.1$ (black bar). For larger driving forces, e.g. $f^\ast_\text{d} = 4.0$, the system always achieves 
a laned state with $\phi\geq 0.8$ (blue (light gray) bars). For medium driving forces, the distribution of $\phi$ yields two peaks (red (dark gray) bars). 
Thus, %\deleted{some simulation runs end up in a laned state and some end up in a jammed state and} 
we may call the system bipartite. 
To decide whether the system is bipartite, we calculate the standard deviation $\sigma_{\phi}$ of the parameter $\phi$ found in different 
simulations (see Fig.~\ref{fig:Bipartite} inset). 
We interpret the system as bipartite if the standard deviation exceeds $\sigma_{\phi}\geq 0.05$. 
 
Interestingly, test simulations with laned initial conditions did not undergo a jamming process. 
This %\deleted{suggests that jamming is meta stable and corresponds to a trapping, which} 
fits to our previous interpretation as an arrested phase separation. 
On the time scales accessible by BD simulations jamming is stable, but on larger time-scales, 
jammed states might eventually transform into laned states.  
For completeness we note that jamming has not been observed in HS systems of oppositely 
driven particle species, except in confinement (which is not the case here)~\cite{Helbing2000}. \\

%----Linear Stability----
\subsection{\label{subsec:PhaseBehaviour}Relation to stability in the effective equilibrium system}

\begin{figure}[]
\includegraphics[width=\columnwidth]{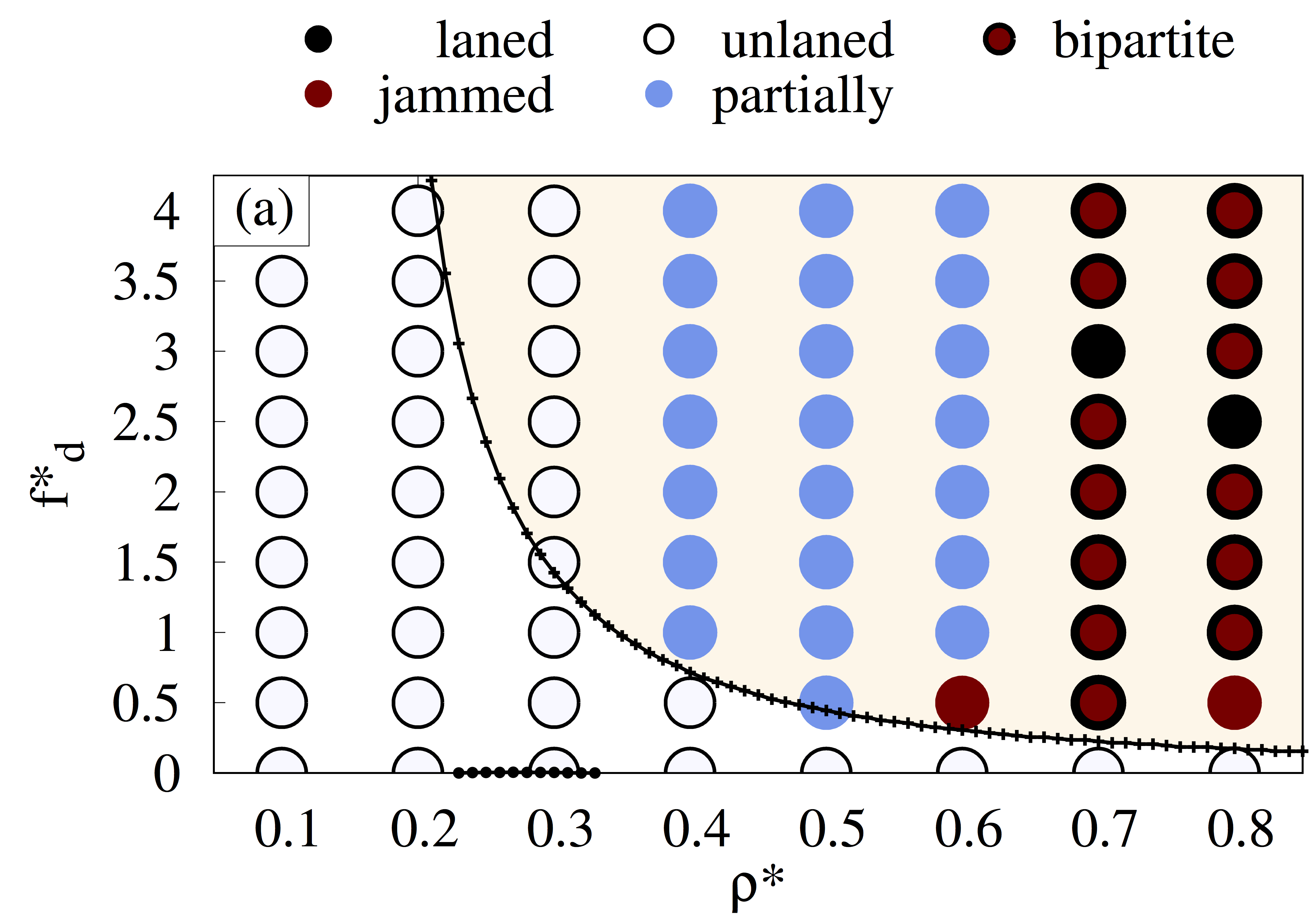}\\
\includegraphics[width=\columnwidth]{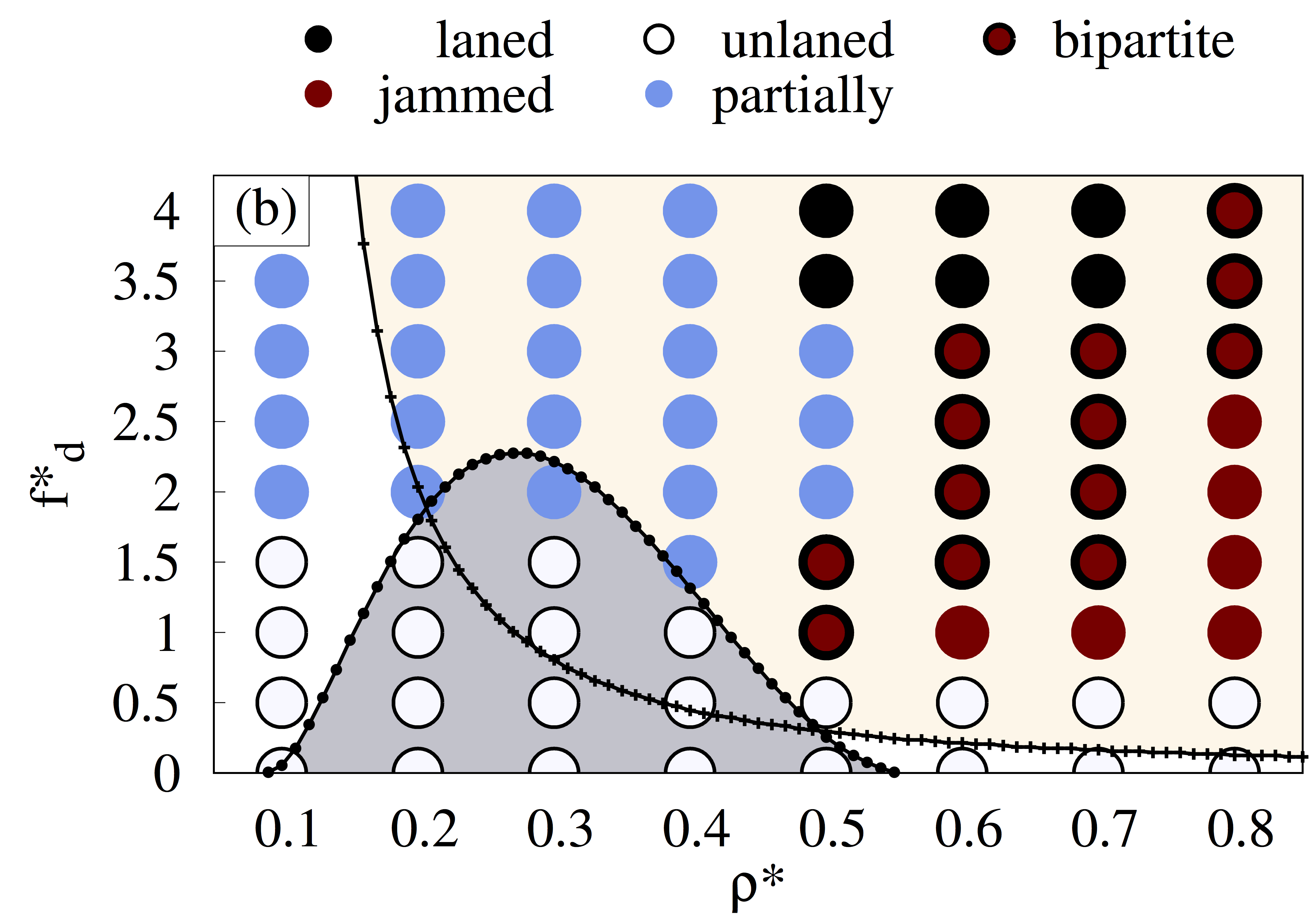}
\caption{\label{fig:StateDiagramm_eps25} State diagram of the system at (a) $\varepsilon^\ast = 2.5$ and (b) $\varepsilon^\ast =4$ 
in the $\varrho^\ast$-$f^\ast_\text{d}$-plane. Circles are results from the simulations and colored according to the four different states as 
discussed above: laned (I, black), jammed (II, red (dark gray)), unlaned (III, white) and partially laned (IV, blue (light gray)). Furthermore we indicated bipartite simulation runs 
by red (dark gray) and black color code. The yellow (light gray) area in the plane denotes the instability due to the demixing of particles calculated from $\gamma(k)_-$ and 
the (dakr) gray area the instability due to the liquid-gas phase separation calculated from $\gamma(k)_+$.}
\end{figure}
In this final section, we relate our findings to those of a 
linear stability analysis based on section \ref{subsec:MethodsDDFT}. The underlying idea is that the onset of laning corresponds to an instability of the homogeneous, fully mixed state. 
In the framework of DDFT, such an instability is indicated if one of the quantities $\gamma_{\pm}(k)$ [see Eq.~(\ref{eqn:Gamma})] 
becomes positive for an arbitrary value of $k$. 
The index '$+$' or '$-$' then indicates the character of the instability.
Our numerical calculations reveal that there are indeed parameter regions where $\gamma_{+}(k)$ or $\gamma_{-}(k)$ become positive.
This always occurs at very small values of $k$ ($k\sigma<10^{-5}$), indicating long-wavelength instabilities.
A positive sign of $\gamma_+$ can thus be interpreted as an instability against (system-spanning) fluctuations of the total density, i.e. a precursor for condensation.
This instability coincides with the so-called spinodal related to
the change of sign  of the quantity $A_+(k\to 0)$ defined below Eq.~(\ref{eqn:Gamma}). 
The second instability, denoted by '$-$', emerges due to a system-spanning demixing of particle species. 
We stress that in the framework of our stability analysis, we investigate an effective model system where the interaction [see Eq.~(\ref{eq:effectivepotential})] between different particle species is corrected by the driving force.
This differs from the analysis in~\cite{kogler1}, where we searched for condensation of the true equilibrium system ($f_d=0$).

In Fig.~\ref{fig:StateDiagramm_eps25} (a) we have plotted the state diagram of the system at  $\varepsilon^\ast=2.5$
in the plane spanned by density and driving force (see also Fig.~\ref{fig:StateDiagramm1}) together with the results of our stability analysis.
The latter shows that a condensation instability ($\gamma_+>0$) only occurs in the equilibrium system ($f^\ast_d=0$) at densities $ \rho^\ast$ in the interval $[0.23,0.33]$.
This interval lies inside the binodal found in Monte Carlo simulations of the 2D one-component LJ-fluid~\cite{smit_frenkel:LJ2D}(recall that the critical coupling strength of the 2D Lennard-Jones Fluid is $\varepsilon^\ast_c=2.17$).
Moreover, visual inspection of simulation
snapshots reveals the existence of large clusters (composed of both particle species) in the system.
Thus,  our stability analysis is consistent with the previously found equilibrium phase behavior of the (effectively one-component) system. 

At finite driving forces ($f_d^\ast>0$) the  effective system transforms into a true 
binary mixture [see Eq.~(\ref{eq:effectivepotential})]. From Fig.~\ref{fig:StateDiagramm_eps25}(a) we see that, in this case, the condensation transition becomes suppressed. 
However, the two-component character of the driven system allows for a second instability, that is, demixing. 
We marked the area of this demixing instability in yellow (light gray).
Interestingly, all laned (and jammed) states identified from simulations are enclosed by the corresponding line of instability.
%Most interestingly, the demixing transitionand shows no relation to the laning transition. 

We now turn to the case $\varepsilon^\ast = 4.0$, see Fig.~\ref{fig:StateDiagramm_eps25}(b).
Again, the demixing instability line encloses laned and jammed states (except for very small densities, where laning is not properly detected).  
In this sense, we find a semi-quantitative agreement between simulations and theoretical predictions of lane formation. 
Another feature apparent from Fig.~\ref{fig:StateDiagramm_eps25}(b) is that the condensation instability now exists also at finite driving forces $f_d^\ast>0$ 
(area marked in (dark) gray).
Furthermore, there exists a region where the curves indicating demixing and condensation overlap.
In parallel, simulations show that lane formation is suppressed in this overlap region. 
This indicates that demixing (laning) and condensation indeed compete in this range of density.

Our conceptional understanding of this subtle interplay between condensation and demixing is as follows:  
The equilibrium LJ fluid ($f_d^\ast=0$) undergoes condensation for temperatures below its critical coupling strength $\varepsilon_{c}^\ast=2.17$~\cite{smit_frenkel:LJ2D,kogler1}.
At coupling strengths $\varepsilon^\ast \gtrsim \varepsilon^\ast_c$, the \textit{driven} system easily forms lanes~\cite{kogler1}.
Within our effective model, the reason is that the interaction between particles of same species equals the true LJ interaction, which is more attractive
than the interaction between particles of opposite species (see Sec.~\ref{subsec:DDFT}). 
Hence we observe, at $\varepsilon^\ast \gtrsim \varepsilon_c^\ast$, a 'local' condensation of particles of the \textit{same} species, but not necessarily an overall
condensation involving both species.
This translates into a demixing of the driven system ($\gamma (k)_-$ instability) in the framework of our stability analysis.

Further increase of the coupling strength to, e.g., $\varepsilon^\ast=4$, leads to strongly attractive interaction between \textit{all} species.
This allows for an overall condensation, that is, an $\gamma(k)_+$ instability even for $f_d^\ast>0$.
This  condensation competes with the demixing process and suppresses laning, because now particles of opposite species (being members of different lanes) tend to aggregate.
We note, however, that the $\gamma(k)_+$ instability is not the reason for the jamming transition at higher densities.
There the system is in principle demixed, but along the direction perpendicular to the drive (see Sec.~\ref{subsec:StructureAnalysis4}).\\
Importantly, the partially laned states found at the very small density $\rho^\ast=0.1$ at $\varepsilon^\ast=4.0$ [see Fig.~\ref{fig:StateDiagramm_eps25}(b)] are not laned according to the linear stability analysis.
We suspect that our characterization of partially laned states most probably fails there.

Finally, for very dense systems the linear stability analysis predicts laning 
already for very low driving forces. This is not mirrored by the BD simulations.
However, it is reasonable to assume 
that due to the very strong attraction strengths, the dynamics of lane formation 
is slowed down to an extent that the simulation times required to see lane formation 
are just not accessible.

\section{Conclusion}
In this paper we have studied a nonequilibrium version of a binary LJ fluid, 
where the two particle species differ in the direction along which they are driven by an external force. The equilibrium LJ fluid can be considered as a generic model fluid describing
attractive interactions between the particles and condensation phase transitions. Thus, we here investigate a representative model allowing to unravel
the interplay of condensation and laning, a typical nonequilibrium transition also occuring in driven repulsive systems.

By extensive BD simulations for a large range of densities and driving forces 
we have identified four different states, namely perfectly laned, partially laned, jammed and unlaned states. These states have been characterized by the laning 
order parameter $\phi$, the mean extension $\left<l_y\right>$ of the clusters in the direction of the driving force, and by other structural measures.
For comparison, the purely repulsive driven hard sphere systems only displays unlaned and perfectly laned states \cite{finite_size:lane}. We thus conclude that the attraction
induces new physics in the system.

Our simulation results for the parameter regimes where laning occurs are supported by the results of a linear stability analysis. 
To this end we have employed DDFT for an effective \textit{equilibrium} (non-driven) system, in which the impact of the driving force 
is taken into account via a corrected pair potential between oppositely
driven particles. The latter is characterized by a weakening of the attractive potential well relative to the undriven case.
The effective model thus corresponds to a binary LJ mixture in which the unlike interactions are weaker than those within each species. As a consequence,
the resulting many-particle system can display not only overall condensation, but also demixing transitions, i.e., the formation of phases with different composition. Our picture is
that the demixing in the effective system (which goes along with a spatial separation of the two species) 
represents the analog of the laning transition in the original driven system.

The results of the stability analysis then indeed suggest
that driven attractive systems are governed by a competition between demixing (laning) and condensation. 
This competition becomes particularly important at large strength of the attractive interactions, where the
system has a strong tendency to condensate and thus, laning is destabilized.
Clearly, our equilibrium theory is strongly simplified. On top of the underlying idea that there
is some ``free energy'' governing the non-equilibrium system (an idea which has been repeatedly criticized,
see, e.g., Ref.~\cite{loewenagainsteffective}), our effective model is also inaccurate 
in that it only takes into account the
impact of the drive on the particles of different species, but not on those of the same type. Still, 
the results of the stability analysis turned out to be quite consistent with those from the BD simulations (see Fig.~8).

From a more general perspective, we note that attractive interactions in driven colloidal systems occur in many contexts, examples being the  depletion interactions in mixtures of different sizes~\cite{loewenpoly},
but also the anisotropic interactions arising in driven dipolar systems~\cite{kogler1} and between colloids driven through liquid crystals~\cite{LCdriven}. We thus expect our results to be applicable also for more complex systems.
Finally, it would be worth to explore connections to active fluids consisting of self-propelled agents with attractive interactions. Indeed, the interplay of
self-organization in such systems and equilibrium phase separation is currently a very lively field of research~\cite{Schwarz-Linek,Redner2013_1,Redner2013_2,Bechinger2013}.\\
%%******End of content*****%%

\begin{acknowledgments}
We thank the DFG for financial support via the International Research Training Group IRTG 1524.
\end{acknowledgments}

\section{Appendix}
In this appendix we give further information on the derivation of Eq.~(\ref{eqn:gamma}), which is the central equation for the linear stability analysis.\\
Inserting the expansion of the functional derivative of the free energy functional $\mathcal F$ [Eq.~(\ref{eqn:Expansion})] into Eq.~(\ref{eqn:DDFT}) results into two coupled differential equations. For the '+'-species we obtain
\begin{align}\label{DGL-DDFT1}
\begin{split}
&\frac{\partial \Delta \varrho(\mathbf r,t)}{\partial t} = \Gamma \nabla \cdot \nabla \left[ \Delta \varrho(\mathbf r,t)\right. \\
&\qquad\left.+\varrho_+^0 \int d\mathbf r' \left.\frac{\delta^2\mathcal F_{ex}[\varrho_+,\varrho_-]}{\delta\varrho_+\delta\varrho_+}\right|_{\varrho_+^0,\varrho_-^0} \Delta \varrho(\mathbf r,t)\right. \\
&\qquad \left.+\psi\varrho_+^0\int d\mathbf r' \left.\frac{\delta^2\mathcal F_{ex}[\varrho_+,\varrho_-]}{\delta\varrho_+\delta\varrho_-}\right|_{\varrho_+^0,\varrho_-^0}\Delta \varrho(\mathbf r,t)\right]~,
\end{split}
\end{align}
and for the '-'-species
\begin{align}\label{DGL-DDFT2}
\begin{split}
&\psi\frac{\partial \Delta \varrho(\mathbf r,t)}{\partial t} = \Gamma \nabla \cdot \nabla \left[\psi \Delta \varrho(\mathbf r,t) \right.\\
&\qquad \left.+\psi\varrho_-^0 \int d\mathbf r' \left.\frac{\delta^2\mathcal F_{ex}[\varrho_+,\varrho_-]}{\delta\varrho_-\delta\varrho_-}\right|_{\varrho_+^0,\varrho_-^0} \Delta \varrho(\mathbf r,t)\right. \\
&\qquad \left.+\varrho_-^0\int d\mathbf r' \left.\frac{\delta^2\mathcal F_{ex}[\varrho_+,\varrho_-]}{\delta\varrho_-\delta\varrho_+}\right|_{\varrho_+^0,\varrho_-^0}\Delta \varrho(\mathbf r,t)\right]~,
\end{split}
\end{align}
where $\Gamma = D/k_\text{B}T$.\\
Performing a Fourier transform of Eqs.~(\ref{DGL-DDFT1}) and (\ref{DGL-DDFT2}) with respect to the position coordinate and introducing the direct correlation functions [see Eq.~(\ref{DCF-free})] yields Eqs.~(\ref{eqn:momentumSpace1}) and (\ref{eqn:momentumSpace2}). 
The latter can be rewritten in a matrix representation in order to find the solutions of $\gamma(k)$ which satisfys both 
equations simultaneously. That is,  
\begin{equation}\label{eqn:Matrix}
\gamma(k) \left(\begin{array}{c} 1 \\ \psi \end{array}\right) = \underline{\underline{M}}\cdot \underline{\underline{G}}\left( \begin{array}{c} 1 \\ \psi \end{array} \right)~,
\end{equation}
with the $2\times 2$ matrices 
\begin{align}
\underline{\underline{M}} = &\left(\begin{array}{cc} -k^2\Gamma & 0 \\ 0 & -k^2\Gamma \end{array}\right) ~, \\
\underline{\underline{G}} =& \left(\begin{array}{cc}1-\tilde c_{++}^{(2)}\varrho_+^0 & -\tilde c_{+-}^{(2)}\varrho_+^0 \\ \\
-\tilde c_{-+}^{(2)}\varrho_-^0 & 1-\tilde c_{--}^{(2)}\varrho_-^0\end{array}\right)~.
\end{align}
Since $\underline{\underline{M}}$ is diagonal and the diagonal elements are non-zero, the inverse $\underline{\underline{M}}^{-1}$ exists and the solutions of Eq.~(\ref{eqn:Matrix}) reads
\begin{align}\label{eqn:gamma_appendix}
\begin{split}
\gamma(k) =& \frac{Tr\left(\underline{\underline M}\cdot \underline{\underline G}\right)}{2}\pm\sqrt{\frac{Tr\left(\underline{\underline M}\cdot \underline{\underline G}\right)^2}{4}-det \left(\underline{\underline M}\cdot \underline{\underline G}\right)} \\
=& \frac{k^2\Gamma}{2}\left(\tilde c_{--}^{(2)}\varrho_-^0+\tilde c_{++}^{(2)}\varrho_+^0 -2\right)\\
&\pm \frac{k^2\Gamma}{2}\left[\left(\tilde c_{++}^{(2)}\varrho_+^0\right)^2+\left(\tilde c_{--}^{(2)}\varrho_-^0\right)^2\right.\\ 
&\left.+4\tilde c_{+-}^{(2)}\tilde c_{-+}^{(2)}\varrho_+^0\varrho_-^0 -2\tilde c_{--}^{(2)}\tilde c_{++}^{(2)}\varrho_+^0\varrho_-^0\right]^{1/2}~,
\end{split}
\end{align}
which is identical to Eq.~(\ref{eqn:gamma}).
\end{document}